\documentstyle[epsf]{article}

\def \a{\alpha}
\def \b{\beta}
\def \g{\gamma}

\def \z{\zeta}

\def \k{\kappa}

\def \m{\mu}
\def \n{\nu}

\def \vp{\varpi}
\def \r{\rho}

\def \ph{\phi}
\def \vp{\varphi}
\def \c{\chi}
\def \ps{\psi}
\def \om{\omega}

\def \La{\Lambda}

\def \Ps{\Psi}

\def \la#1{\label{#1}}
\def \ift{\infty}
\def \le{\left}
\def \ri{\right}

\def \ti#1{\tilde{#1}}
\def \lb{\lbrack}
\def \rb{\rbrack}

\def \rar{\rightarrow}

\def \ld{\ldots}
\def \cd{\cdots}
\def \nn{\nonumber}

\newcommand \beq{\begin{eqnarray}}
\newcommand \eeq{\end{eqnarray}}
\newcommand \bea{\begin{eqnarray*}}
\newcommand \eea{\end{eqnarray*}}
\newcommand \ben{\begin{enumerate}}
\newcommand \een{\end{enumerate}}
\newcommand \ba{\begin{array}}
\newcommand \ea{\end{array}}

\def \Tr{{\rm Tr}}

\begin{document}

\begin{center}
   {\LARGE\bf Noncommutative Probability,} \\
   \vspace{.1cm}
   {\LARGE\bf Matrix Models, and} \\
   \vspace{.1cm}
   {\LARGE\bf Quantum Orbifold Geometry} \\
   \vspace{.7cm}
   {\large {\bf C.-W. H. Lee}$^{a,b,}$\footnote
           {e-mail address: h11lee@math.uwaterloo.ca}} \\
   \vspace{.7cm}
   $^a$ {\it Department of Pure Mathematics, Faculty of Mathematics, 
   University of Waterloo, Waterloo, Ontario, Canada, N2L 3G1.} \\
   $^b$ {\it Department of Physics, Faculty of Science, University of 
   Waterloo, Waterloo, Ontario, Canada, N2L 3G1.} \\
   \vspace{.4cm}
   {\large March 11, 2003} \\
   \vspace{.7cm}
   {\large\bf Abstract}
\end{center}

\noindent
Inspired by the intimate relationship between Voiculescu's noncommutative
probability theory (of type~A) and large-$N$ matrix models in physics, we look 
for physical models related to noncommutative probability theory of type~B.  
These turn out to be fermionic matrix-vector models at the double large-$N$ 
limit.  In the context of string theory, they describe different orbifolded 
string worldsheets with boundaries.  Their critical exponents coincide with 
that of ordinary string worldsheets, but their renormalised tree-level 
one-boundary amplitudes differ.

\vspace{.5cm}

\begin{flushleft}
\end{flushleft}
\pagebreak

\section{Introduction}
\la{s1}

The interplay between noncommutative probability and large-$N$ matrix models 
is a wonderful example of how mathematics and physics mutually benefit from 
each other.  Noncommutative probability theory (of type~A) was originally 
developed by Voiculescu as a tool to analyse some von Neumann algebras 
\cite{vdn}; this analysis was rendered possible by random matrices.  It can be 
shown that a large class of random matrices is asymptotically free as the 
order of the matrices is taken to infinity.  Hence it constitutes a major 
example of noncommutative probability.

Nonetheless, random matrices show up in physics as well.  An important 
example is Yang--Mills theory.  The ``infinite-dimensional'' Yang--Mills 
matrix field was dubbed the master field \cite{witten}.  The master field 
enjoys greatly simplifying features and is believed to hold the key to a 
deeper understanding of Yang--Mills theory.  Among other examples are quantum 
gravity and string theory.  The large-$N$ expansion of matrix models was found 
out to be a genus expansion \cite{thooft}; the dual of the Feynman diagrams of 
the leading term may be treated as discretised string worldsheets with the
topology of a sphere \cite{david85, kkm, adf}.  The commonality of random 
matrices in mathematics and physics triggers an interest in the relationship 
between noncommutative probability theory of type~A and large-$N$ matrix 
models \cite{douglas, gg, douglasli, aakv, engelhardt, aakr}.  It is now 
possible to think of notions of noncommutative probability in physical terms 
and describe the master field in algebraic terms.

However, mathematicians have recently realised that other noncommutative 
probability spaces exist.  This is achieved by a reformulation of 
noncommutative probability theory of type~A in terms of non-crossing 
partitions of type~A \cite{speicher}, which, in turn, is related to the 
M\"{o}bius inversion theory in the lattices of non-crossing partitions 
\cite{kreweras}.  Then the fact that other types of non-crossing partitions 
exist \cite{reiner} prompted an attempt to develop a ``noncommutative 
probability theory of type~B'' from non-crossing partitions of type~B 
\cite{bgn}.  Given the intimate relation between noncommutative probability 
theory of type~A and large-$N$ matrix models, naturally we would like to ask 
if there exist large-$N$ matrix models ``of type~B''.

Yes, such matrix models exist; they are fermionic matrix-vector models in the
double large-$N$ limit.  Fermionic matrix (vector) models \cite{mz} are models 
in which the matrix entries (vector components) are Grassmann numbers.  Their 
convergent behavior is, in general, better than their bosonic counterparts.  
Physically, they describe two-dimensional quantum gravity, random polymers 
\cite{akm, mss, ss}, or induced gauge theory \cite{km}.  The basic ingredients 
of the models in this article are $N_v$-dimensional vectors of $N_m \times 
N_m$ matrices of Grassmann numbers; these are thus simultaneously matrices and 
vectors.  We will take $N_v$ to infinity first and $N_m$ to infinity 
afterwards.  Then the dual of the Feynman diagrams of these models describes 
orbifolded string worldsheets in string theory.

We will deal with the physical aspect of fermionic matrix-vector models in 
this article in detail; the more mathematical aspect will be discussed in 
future works.

Here is a synopsis of this article.  We will review the relationship between
noncommutative probability theory of type~A and large-$N$ matrix models in 
Section~\ref{s2}.  Then we will introduce non-crossing partitions of type~B 
and a class of fermionic matrix-vector models which generates these partitions 
in Section~\ref{s3}.  In Section~\ref{s4}, we will identify these partitions
as random surfaces of a bounded region of the orbifold ${\bf R}^2 / {\bf Z}_2 
\times {\bf Z}_2$ and compute the full Green functions (moments) at the double 
large-$N$ limit.  We will also show by means of the Schwinger--Dyson equation 
that the renormalised tree-level one-boundary amplitude and critical exponent 
associated with the string susceptibility at the continuum limit are identical 
to those of ordinary quantum gravity.  In Section~\ref{s5}, we will construct 
a different class of fermionic matrix-vector models which generates random 
surfaces of a bounded region of the quantum orbifold ${\bf R}^2 / {\bf Z}_2$.  
We will show that its critical exponent is also the same as that of ordinary 
quantum gravity, but it has a different one-boundary amplitude.  Finally, we 
will indicate possible extensions of this work in Section~\ref{s6}.

\section{Noncommutative Probability \& Large-$N$ Matrix Models}
\la{s2}

A {\em noncommutative probability space} (of type~A) was originally defined
by D. Voiculescu \cite{vdn} as an ordered pair $({\cal A}, \vp)$, where
${\cal A}$ is a complex unital C$^*$-algebra and $\vp : {\cal A} \rar {\bf C}$ 
a positive linear functional satisfying the normalisation condition that 
$\vp(1) = 1$.  Each element $a$ of ${\cal A}$ may be considered as a random 
variable and $\vp(a)$ its expectation value.

A useful notion of noncommutative probability theory is the non-crossing
cumulant \cite{speicher}.  To define what it is, let us digress for a moment
and introduce a few auxiliary notions first.  Consider a finite set $F$ of
integers.  Let $p$ be a partition of $F$.  We write $n_1 \simeq_p n_2$ if
$n_1$ and $n_2$ are numbers of $F$ such that they are in the same block (or
cell) of the partition $p$.  Furthermore, $p$ is a {\em non-crossing partition 
of type~A} if $n_1 \simeq_p n_3$ and $n_2 \simeq_p n_4$ imply that $n_2 \simeq 
n_3$ for any four integers $n_1$, $n_2$, $n_3$, and $n_4$ in $F$ such that 
$n_1 < n_2 < n_3 < n_4$.  An example of a non-crossing partition of type~A is 
illustrated in Fig.\ref{f1}.  The set of all non-crossing partitions of type~A 
of $F$ is denoted as $NC^{(A)} (F)$.  If $F = \{ 1, 2, \ld, n \}$, we will 
abbreviate $NC^{(A)} (F)$ as $NC^{(A)} (n)$ as there will be no danger of 
confusion.

\begin{figure}[ht]
\epsfxsize=3in
\centerline{\epsfbox{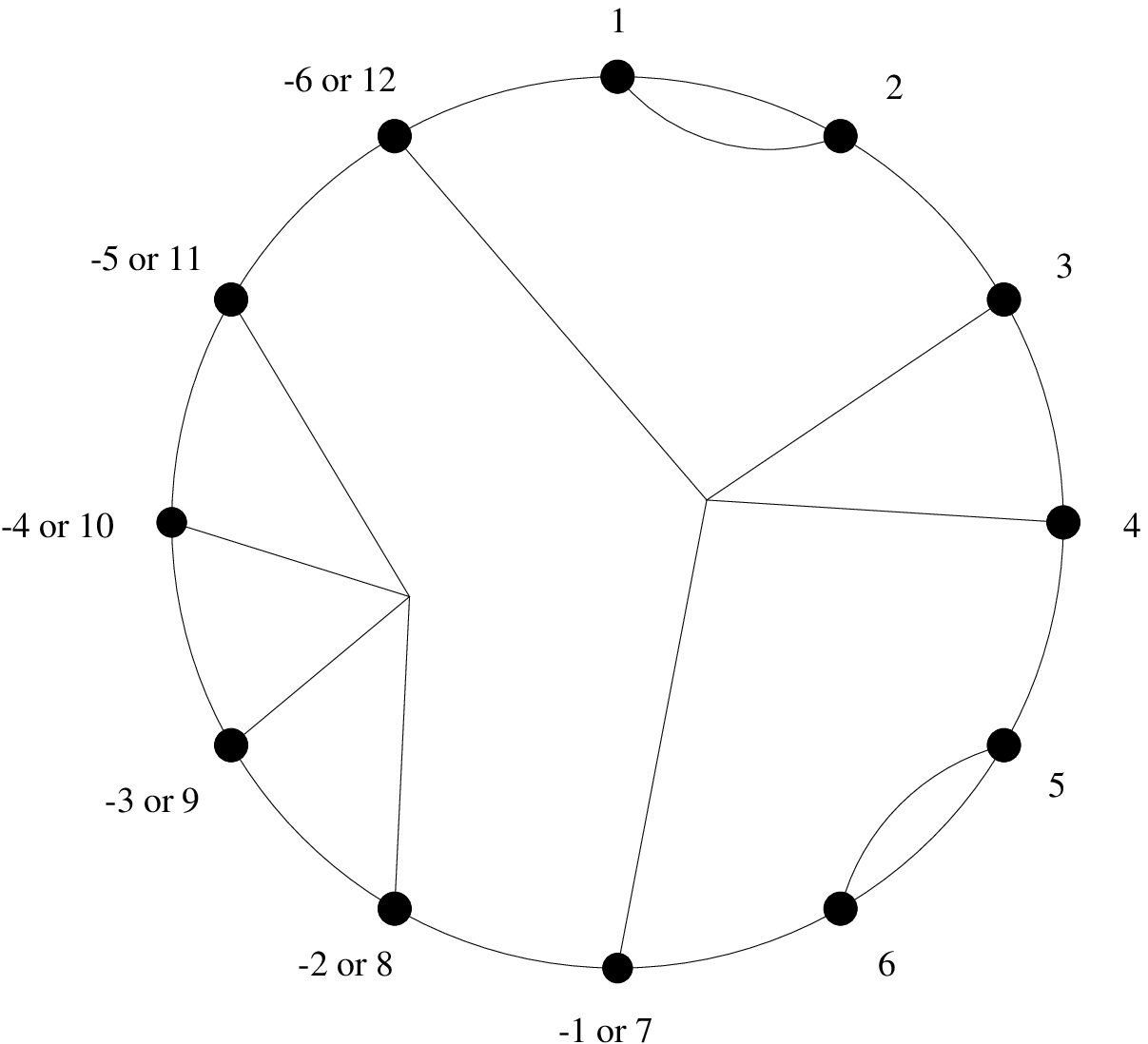}}
\caption{\em A non-crossing partition of type~A but not of type~B.  The
partition of the numbers 1, 2, \ld, 12 into the blocks $\{ 1, 2 \}$, $\{ 3, 4, 
7, 12 \}$, $\{ 5, 6 \}$, and $\{ 8, 9, 10, 11 \}$ is the non-crossing 
partition of type~A, but the (same) partition of the numbers 1, 2, \ld, 6, -1, 
-2, \ld, -6 into the blocks $\{ 1, 2 \}$, $\{ 3, 4, -1, -6 \}$, $\{ 5, 6 \}$, 
and $\{ -2, -3, -4, -5 \}$ is not a non-crossing partition of type~B.  This 
may also be regarded as a planar Feynman diagram with each intersection point 
and the lines connected to it inside the great circle representing a connected 
Green function.}
\la{f1}
\end{figure}

A {\em non-crossing cumulant} is a multilinear functional $\k_n : {\cal A}^n
\rar {\bf C}$, where $n$ is any positive integer, such that
\beq
   \vp(a_1 a_2 \cd a_k) = \sum_{p \in NC^{(A)} (k)}
   \prod_{\mbox{$F$ is a block of $p$}} \k_{\mbox{card($F$)}}
   ( (a_1, a_2, \ld, a_k) | F).
\la{2.1}
\eeq
In this formula, if $F = \{j_1, j_2, \ld, j_m\}$ such that $j_1 < j_2 < \ld <
j_m$, then card$F$, the cardinality of $F$, is $m$, and
\[ (a_1, a_2, \ld, a_k) | F := (a_{j_1}, a_{j_2}, \ld, a_{j_m}) \in 
   {\cal A}^m. \]
We may apply Eq.(\ref{2.1}) recursively to obtain the expressions of all
non-crossing cumulants in terms of $\vp$.  For example,
\beq
   \k_1 (a_1) & = & \vp (a_1), \nn \\
   \k_2 (a_1, a_2) & = & \vp (a_1 a_2) - \vp(a_1) \vp(a_2), \nn \\
   \k_3 (a_1, a_2, a_3) & = & \vp (a_1 a_2 a_3) - \vp(a_1) \vp(a_2 a_3)
   - \vp(a_2) \vp(a_1 a_3) \nn \\
   & & - \vp(a_3) \vp(a_1 a_2) + 2 \vp(a_1) \vp(a_2) \vp(a_3),
\la{2.2}
\eeq
and so on.  The significance of non-crossing cumulants lies in the fact that
they provide a particularly simple criterion to determine whether a family of
unital subalgebras of ${\cal A}$ is free and that the generating series of
the non-crossing cumulants, called the R-transform, display simplifying
properties in freeness computations \cite{vdn}.

Random matrix models in the large-$N$ limit furnish a nice physical example
of noncommutative probability theory of type~A \cite{gg}.  In the simplest 
case, the random variables are, intuitively speaking, generated by a Hermitian
random matrix $M$ of infinite order, which is alternatively known as the 
master field.  The partition function $Z_b$ of this random matrix model takes 
the form
\[ Z_b = \int dM \exp (- N \Tr V(M)), \]
where $N$ is the order of the matrix $M$ and will eventually be taken to
infinity, $V(M)$ is a polynomial of $M$, and $dM$ is the corresponding 
measure.  The expectation value of the random variable $M^n$ is
\beq
   \ti{\ph}(n/2) := \lim_{N \rar \ift} \frac{1}{Z_b} \int dM \frac{1}{N}
   \Tr M^n \exp (- N \Tr V(M)).
\la{2.3}
\eeq
$\ti{\ph} (n/2)$ is called a {\em full Green function} in physics literature.
The non-crossing cumulant $\k_n (M, M, \ld, M)$ is nothing but the connected 
Green function
\[ \lim_{N \rar \ift} \frac{1}{N} \langle \Tr M^n \rangle_c, \]
and a non-crossing partition of type~A may be represented by a planar Feynman
diagram (Fig.\ref{f1}).  Formulae (\ref{2.2}) are the expressions of connected
Green functions in terms of full Green functions familiar to physicists.

\section{A Fermionic Matrix-Vector Model}
\la{s3}

There are other types of non-crossing partitions which are of interest in
mathematics \cite{reiner}.  For instance, consider the totally ordered set
\[ \lb \pm n \rb := \{ 1 < 2 < \cd < n < -1 < -2 < \cd < -n \}. \]
It is obviously isomorphic to the set
\[ \lb 2n \rb := \{ 1 < 2 < \cd 2n \}. \]
Moroever, define the {\em inversion map} to be a map which maps the integer
$a$ to $-a$.  {\em A non-crossing partition of type~B} of $\lb \pm n \rb$ is a
non-crossing partition of type~A of $\lb 2n \rb$ such that the partition is
invariant under the inversion map. Not all non-crossing partitions of type~A
are of type~B as well; this can be seen in Fig.\ref{f1}.  The set of all
non-crossing partitions of type~B of $\lb \pm n \rb$ is denoted as $NC^{(B)}
(n)$.  A cell of a non-crossing partition of type~B is called a 
{\em zero-block} if it is invariant under the inversion map; otherwise, it is
called a {\em non-zero-block}.  It can be straightforwardly shown that there
is at most one zero-block in every non-crossing partition of type~B.  A 
partition with a zero-block is illustrated in Fig.\ref{f2}.

\begin{figure}[ht]
\epsfxsize=3in
\centerline{\epsfbox{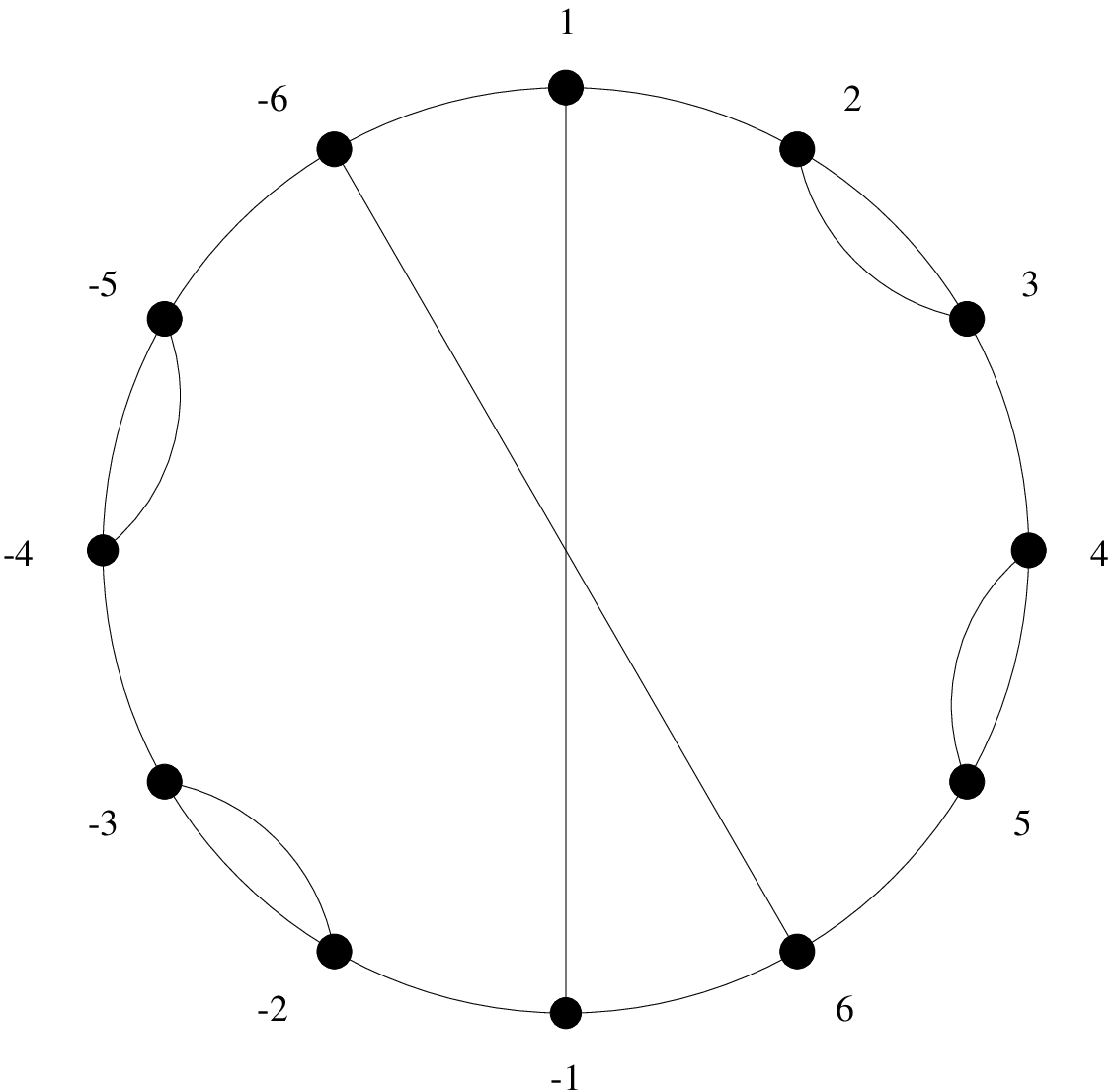}}
\caption{\em A non-crossing partition of type~B.  The partition of the numbers
1, 2, \ld, 6, -1, -2, \ld, -6 into the blocks $\{ 1, 6, -1, -6 \}$, $\{ 2, 3 
\}$, $\{ 4, 5 \}$, $\{ -2, -3 \}$, and $\{ -4, -5 \}$ is a non-crossing 
partition of type~B.  $\{ 1, 6, -1, -6 \}$ is a zero-block, and all others are
non-zero-blocks.}
\la{f2}
\end{figure}

Associated with non-crossing partitions of type~B is a nascent development of
a noncommutative probability theory of type~B \cite{bgn} in which the
non-crossing cumulants are defined in terms of non-crossing partitions of
type~B.  Given the intimate relationship between noncommutative probability
theory and matrix models, it is worthwhile to search for matrix models which
are physical realisation of noncommutative probability theory of type~B.  It 
turns out that there is such a class of matrix models --- the fermionic 
matrix-vector models in the double large-$N$ limit.

A fermionic matrix-vector model is built out of an $N_v$-dimensional vector of 
matrices $\Ps_1$, $\Ps_2$, \ld, and $\Ps_{N_v}$, and another $N_v$-dimensional 
vector of matrices $\bar{\Ps}_1$, $\bar{\Ps}_2$, \ld, and $\bar{\Ps}_{N_v}$.  
Each of these matrices is of order $N_m$.  Every matrix entry is a Grassmann 
number.  The $i$, $j$-th matrix entry of the matrices $\Ps_{\m}$ and 
$\bar{\Ps}_{\m}$ are denoted by $\ps_{\m ij}$ and $\bar{\ps}_{\m ij}$, 
respectively.  Any two Grassmann numbers anti-commute with each other.

The action of a family of fermionic matrix-vector models takes the form
\beq
   \lefteqn{S := N_m \sqrt{N_v} \sum_{\m = 1}^{N_v} \Tr \bar{\Ps}_{\m} 
   \Ps_{\m}} \nn \\
   & & + N_m \sum_{n=1}^{\ift} \frac{c_n}{2n}
   \sum_{\m_1, \m_2, \ld, \m_{2n} = 1}^{N_v} \Tr \le\lb \le(
   \bar{\Ps}_{\m_1} \Ps_{\m_2} \bar{\Ps}_{\m_3} \Ps_{\m_4} \cd
   \bar{\Ps}_{\m_{2n-1}} \Ps_{\m_{2n}} \ri)^2 \ri\rb \nn \\
   & & + N_m^2 \sum_{n=2}^{\ift} \frac{g_n}{2n}
   \sum_{\m_1, \m_2, \ld, \m_{2n} = 1}^{N_v} \le\lb \Tr \le(
   \bar{\Ps}_{\m_1} \Ps_{\m_2} \bar{\Ps}_{\m_3} \Ps_{\m_4} \cd
   \bar{\Ps}_{\m_{2n-1}} \Ps_{\m_{2n}} \ri) \ri\rb^2,
\la{3.1}
\eeq
where $c_1$, $c_2$, $c_3$, \ld, and so on and $g_2$, $g_3$, $g_4$, \ld, and so
on are constant complex numbers.  The partition function is given by
\beq
   Z_{N_m, N_v} := \int d\Ps_1 d\bar{\Ps}_1 d\Ps_2 d\bar{\Ps}_2 \cd
   d\Ps_{N_v} d\bar{\Ps}_{N_v} \exp S,
\la{3.2}
\eeq
where the Grassmann integrals are defined as usual in the sense of Brezin
\cite{brezin}.  The full Green functions which are of interest to us take the 
form
\[ G_{N_m, N_v} (n) := \frac{1}{N_m} \sum_{\m_1, \m_2, \ld, \m_{2n} = 1}^{N_v}
   \le\langle \Tr \le\lb \le( \bar{\Ps}_{\m_1} \Ps_{\m_2} \bar{\Ps}_{\m_3}
   \Ps_{\m_4} \cd \bar{\Ps}_{\m_{2n-1}} \Ps_{\m_{2n}} \ri)^2 \ri\rb
   \ri\rangle_S, \]
where the subscript $S$ means that this expectation value is evaluated with
respect to the partition function defined by the action $S$.

The Feynman rules are analogous to those of $U(N)$ gauge theory \cite{thooft}.
Some examples are depicted in Fig.\ref{f3}.  Note that a {\em double-end},
i.e, a pair of ends of a double-line representing a propagator or a double-leg
of a vertex, represents a row index of an $N_m \times N_m$ matrix, a column
index of the same matrix and an index of an $N_v$-dimensional vector.  If the
subscript of a vector index is odd, then the corresponding double-end 
originates from $\bar{\Ps}$; otherwise, it originates from $\Ps$.  For 
instance, the indices $\m_1$ and $\m_3$ in the $c_1$-vertex originates from
$\bar{\Ps}$, whereas the indices $\m_2$ and $\m_4$ originates from $\Ps$.  As
a result, we may connect two double-ends together by a propagator only if the
subscript of the vector index of one of the double-ends is odd and the
subscript of the vector index of the other double-end is even.  For example,
we may use a propagator to connect the double-end of a $c_1$ vertex whose
vector index is $\m_1$ to the double-end of a $g_2$ vertex whose vector index 
is $\n_4$, but we may not use a propagator to connect the same double-end of 
the $c_1$ vertex with the double-end of the $g_2$ vertex whose vector index is 
$\n_3$.  Each of the lines connects a row index to a column index; the 
connections have nothing to do with vector indices. A fermionic loop may 
contribute a factor of $N_m$ or $-N_m$.  The sign is determined by first 
principles; two Feynman diagrams which are topologically equivalent after 
removal of the vector indices may carry opposite signs.

\begin{figure}
\epsfxsize=4in
\epsfysize=6.2in
\centerline{\epsfbox{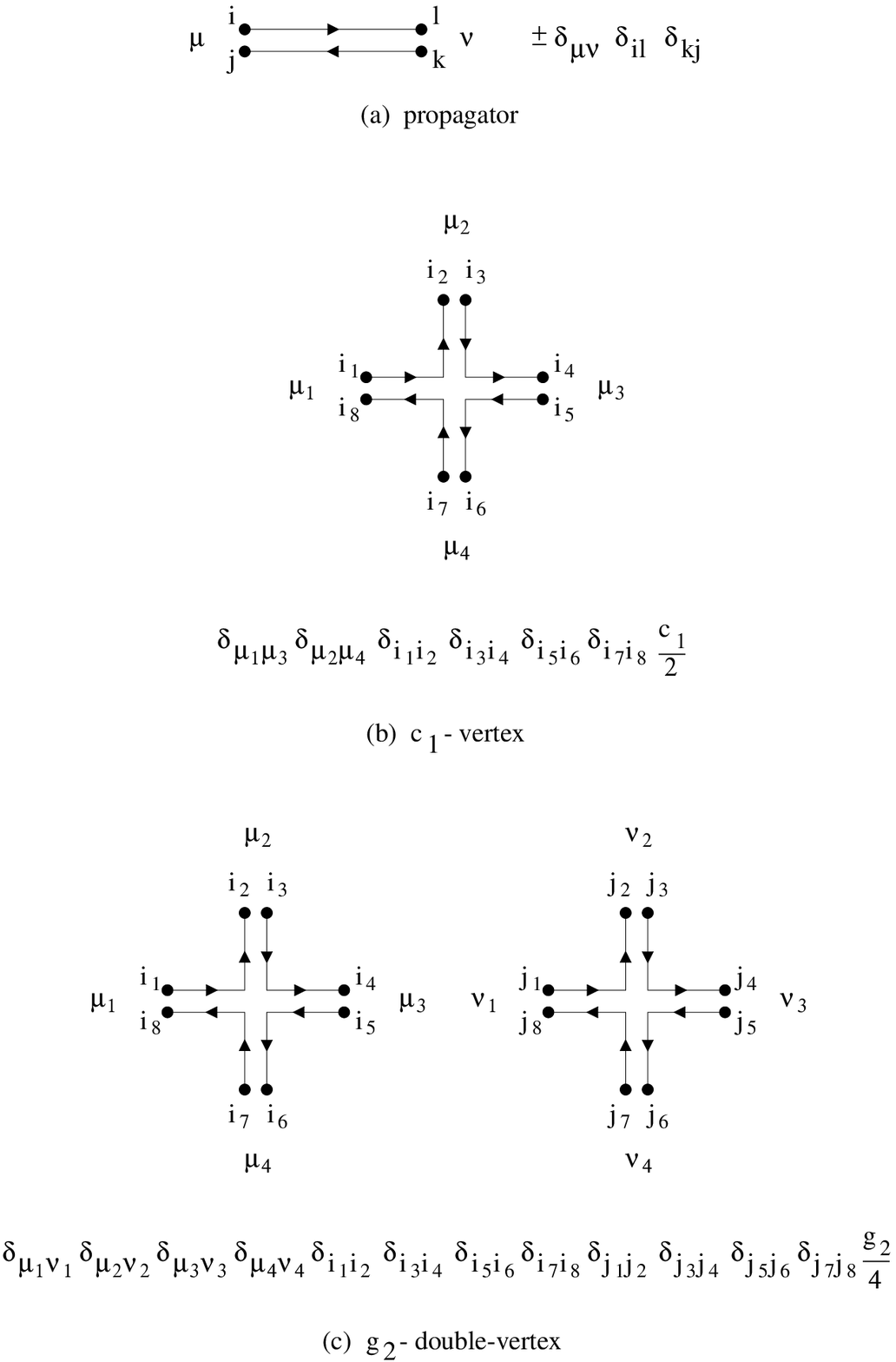}}
\caption{\em Selected Feynman rules for a fermionic matrix-vector model. 
Depicted here are (a) a propagator, (b) a vertex with a coefficient $c_1$, and
(c) a double-vertex with a coefficient $g_2$.  Each pair of solid circles
represents a double-end.  Each fermionic loop contributes, up to a sign, a
factor of $N_m$.}
\la{f3}
\end{figure}

The {\em double large-$N$ limit} is characterised by the following formulae
for the partition function and the Green functions:
\beq
   Z & := & \lim_{N_m \rar \ift} \lim_{N_v \rar \ift} Z_{N_m, N_v},
   \mbox{and} \nn \\
   G(n) & := & \lim_{N_m \rar \ift} \lim_{N_v \rar \ift} G_{N_m, N_v} (n).
\la{3.3}
\eeq
Note that we take $N_v$ to infinity {\em before} taking $N_m$ to infinity;
this ensures that only connected Feynman diagrams will contribute to $G(n)$.
Moreover, only those Feynman diagrams which satisfy the following condition
survive the large-$N_v$ limit:
\begin{quote}
{\em if, in a Feynman diagram, a propagator connects together two double-ends 
whose vector indices are $\m_a$ and $\n_b$, then in the same diagram there 
must be another propagator connecting two other double-ends whose vector 
indices are $\m_a$ and $\n_b$, too.}
\end{quote}
Furthermore, only planar Feynman diagrams which survive the large-$N_v$ limit
contribute to the double large-$N$ limit.  An example of such a Feynman 
diagram is illustrated in Fig.\ref{f4}.  This diagram contributes to
Fig.\ref{f2}, which, as a Feynman diagram, each intersection point and the
lines connected to it inside the great circle represent a connected Green
function, any pair of vector indices $\m_k$ of the sources are abbreviated as 
$k$ and $-k$ for any value of $k$, and the vector indices of the vertices are 
suppressed.

\begin{figure}[ht]
\epsfxsize=4in
\centerline{\epsfbox{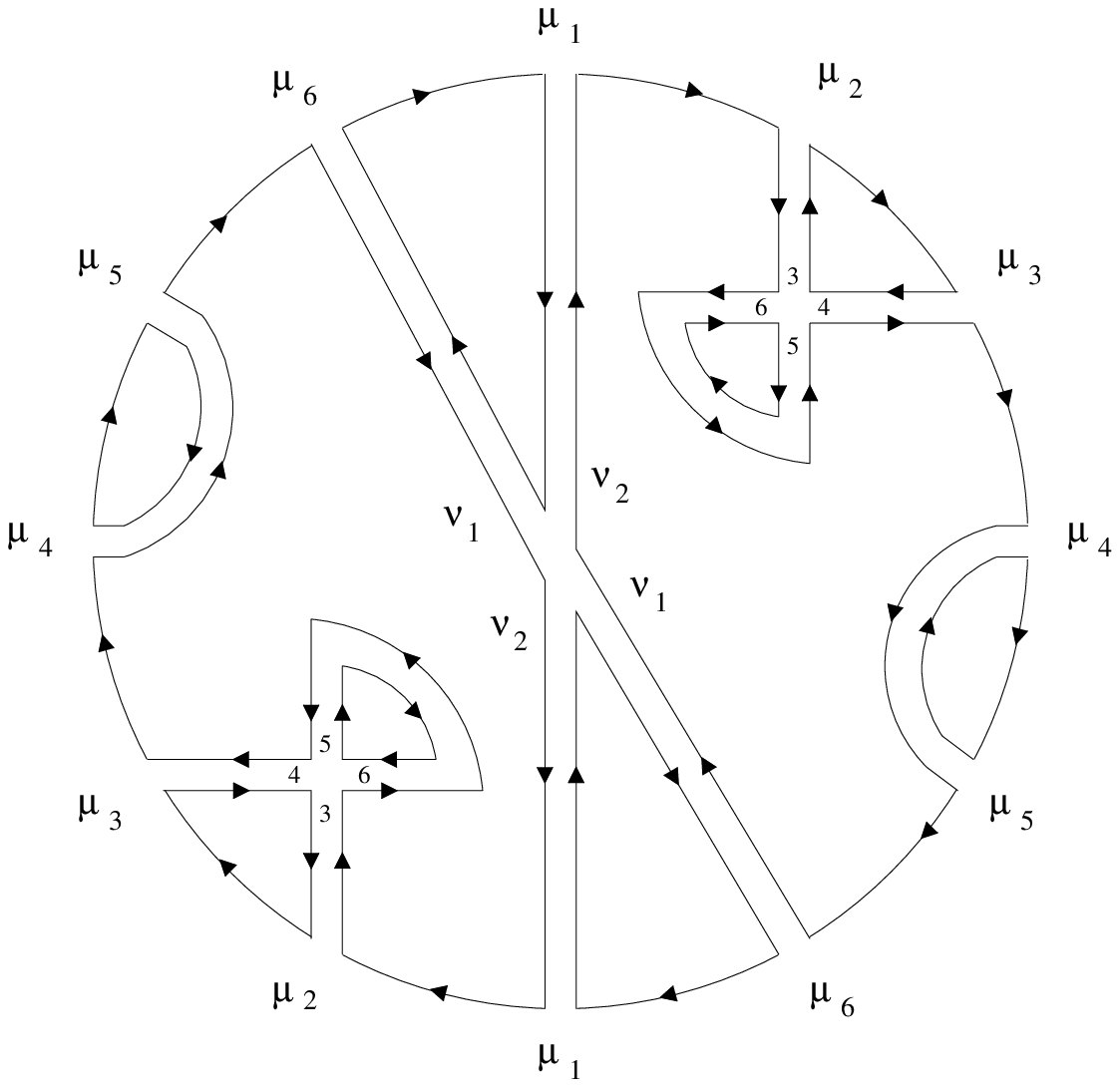}}
\caption{\em A Feynman diagram of $G(3)$.  Only vector indices are shown.  The 
numbers 3, 4, 5, and 6 stand for $\n_3$, $\n_4$, $\n_5$, and $\n_6$, 
respectively.}
\la{f4}
\end{figure}

\section{Quantum Orbifold Geometry}
\la{s4}

It is well known that the dual of the Feynman diagrams of matrix models
associated with noncommutative probability theory of type~A may be regarded as 
discrete random surfaces of two-dimensional quantum gravity.  (See, e.g,
Refs.\cite{dgz} and \cite{adj} and the references therein.)  Does the 
fermionic matrix-vector model play a role in quantum gravity as well?

A clue may be found in Fig.\ref{f4}.  Choose the center of the great circle to
be the origin of a two-dimensional Cartesian coordinate system.  Point the
$y$-axis to the index $\m_1$.  Then the figure respects parity transformation 
$x \rar -x$ and $y \rar -y$ about the origin.  So does its dual.  Though a 
thoroughly rigorous proof is lacking, it is probably true that every Feynman 
diagram of $G(n)$ which contributes to the double large-$N$ limit exhibits the 
same symmetry.  Hence we assert that {\em the dual of the Feynman diagrams of 
$G(n)$ in the double large-$N$ limit may be thought of as quadrangulated 
random surfaces which respect parity transformation about the origin, i.e, 
these are random surfaces of a bounded region of the orbifold ${\bf R}^2 / 
{\bf Z}_2 \times {\bf Z}_2$}.  (c.f. Fig.\ref{f5}.)  

\begin{figure}[ht]
\epsfxsize=4in
\center{\epsfbox{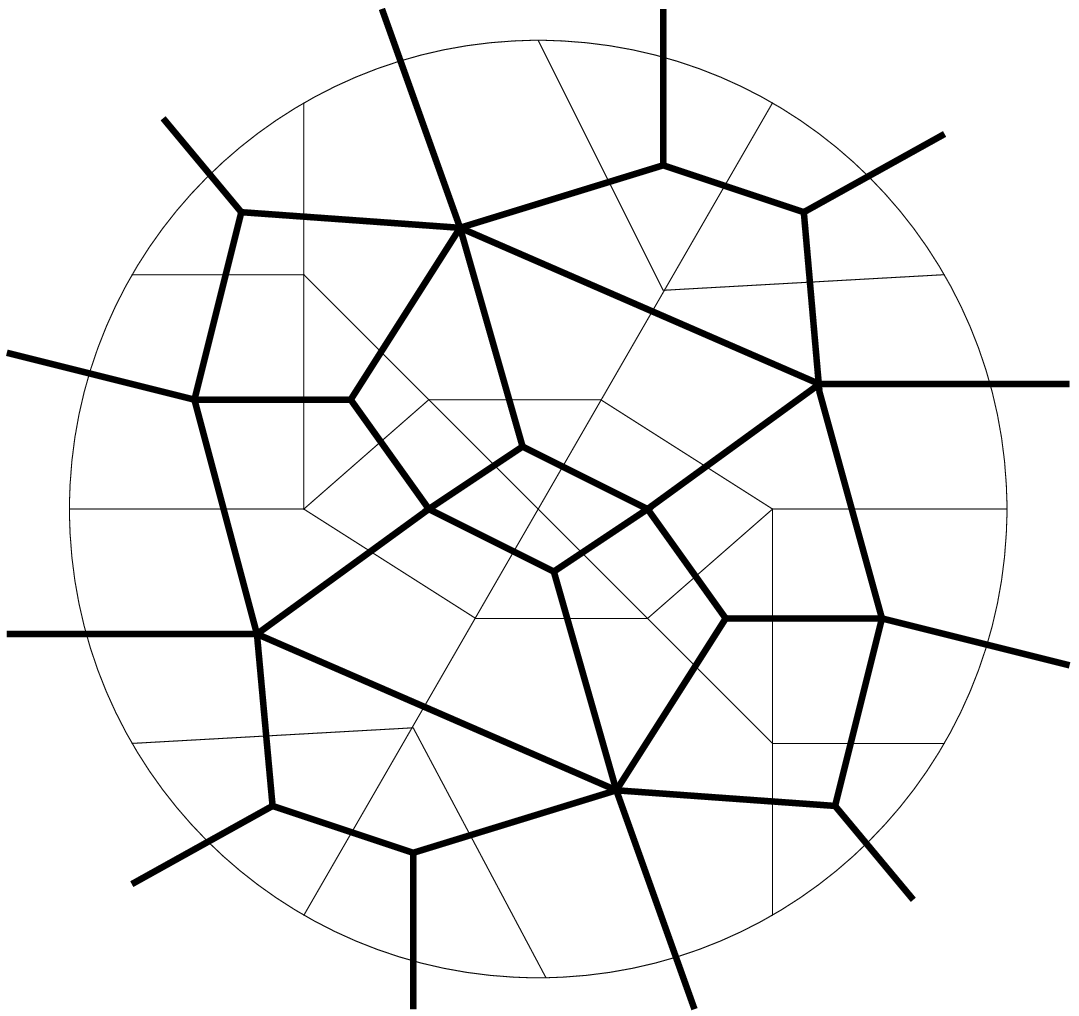}}
\caption{\em A discretized random orbifold surface with a boundary.  The 
quadrilaterals paving the surface are in thick lines, and the thin circle 
forms the boundary.  The dual of this surface, drawn in thin lines, is a 
Feynman diagram of the fermionic matrix-vector model defined in Eq.(\ref{4.1})
in the double large-$N$ limit.  Note that the ``spokes'' protruding from the 
great circle converge to the same point on a spherical surface.}
\la{f5}
\end{figure}

Let us study the full Green functions and critical behavior of the simplest 
fermionic matrix-vector model in which only the coefficients $c = c_1$ and 
$g = g_2$ are non-zero.  From Eq.(\ref{3.1}), the action is
\beq
   S & := & N_m \sqrt{N_v} \sum_{\m = 1}^{N_v} \Tr \bar{\Ps}_{\m} \Ps_{\m}
   + \frac{N_m c}{2} \sum_{\m_1, \m_2 = 1}^{N_v} \Tr \le\lb
   \le( \bar{\Ps}_{\m_1} \Ps_{\m_2} \ri)^2 \ri\rb \nn \\
   & & + \frac{N_m^2 g}{4} \sum_{\m_1, \m_2, \m_3, \m_4 = 1}^{N_v} \le\lb \Tr
   \le( \bar{\Ps}_{\m_1} \Ps_{\m_2} \bar{\Ps}_{\m_3} \Ps_{\m_4} \ri)
\ri\rb^2.
\la{4.1}
\eeq
The full Green functions can be obtained by the Schwinger--Dyson approach.  
Indeed, consider the trivial equation
\beq
   \lefteqn{\lim_{N_m \rar \ift} \lim_{N_v \rar \ift}
   \frac{1}{N_m^2 \sqrt{N_v} Z_{N_m, N_v}} \sum_{i,j=1}^{N_m} 
   \sum_{\b = 1}^{N_v} \int d\Ps_1 d\bar{\Ps}_1 d\Ps_2 d\bar{\Ps}_2 \cd
   d\Ps_{N_v} d\bar{\Ps}_{N_v} } \nn \\
   & & \frac{\partial}{\partial \bar{\Ps}_{\b ij}} \{
   \sum_{\a_1, \a_2, \ld, \a_{2n-1} = 1}^{N_v}
   \le( \bar{\Ps}_{\a_1} \Ps_{\a_2} \cd \bar{\Ps}_{\a_{2n-1}} \Ps_{\b} 
   \bar{\Ps}_{\a_1} \Ps_{\a_2} \cd \bar{\Ps}_{\a_{2n-1}} \ri)_{ij} \nn \\
   & & \exp S \} = 0
\la{4.1.1}
\eeq
where $S$ was given in Eq.(\ref{4.1}) and $n$ is any positive integer.  Since
\bea
   \lefteqn{\lim_{N_m \rar \ift} \lim_{N_v \rar \ift} 
   \frac{1}{N_m^2 \sqrt{N_v}} \sum_{i,j=1}^{N_m} \sum_{\b = 1}^{N_v}  
   \sum_{\a_1, \a_2, \ld, \a_{2n-1}}^{N_v} \le\langle \le( \bar{\Ps}_{\a_1} 
   \Ps_{\a_2} \cd \Ps_{\a_{2k}} \ri. \ri. } \\
   & & \le. \le. \cdot \le( \frac{\partial}{\partial \bar{\Ps}_{\b ij}} 
   \bar{\Ps}_{\a_{2k+1}} \ri) \Ps_{\a_{2k+2}} \cd \bar{\Ps}_{\a_{2n-1}} 
   \Ps_{\b} \bar{\Ps}_{\a_1} \Ps_{\a_2} \cd \bar{\Ps}_{\a_{2n-1}} 
   \ri)_{ij} \ri\rangle_S \\
   & & = \ti{\ph} (k) G (n-k-1), 
\eea
where $\ti{\ph}(n)$ was defined in Eq.(\ref{2.3}) with 
\[ V(M) = \frac{1}{2} M^2 - \frac{g}{4} M^4 \]
and $0 \leq k \leq n-2$, 
\bea
   \lefteqn{\lim_{N_m \rar \ift} \lim_{N_v \rar \ift}
   \frac{1}{N_m^2 \sqrt{N_v}} \sum_{i,j=1}^{N_m} \sum_{\b = 1}^{N_v} 
   \sum_{\a_1, \a_2, \ld, \a_{2n-1}}^{N_v} \le\langle \le(
   \bar{\Ps}_{\a_1} \Ps_{\a_2} \cd \Ps_{\a_{2n-2}} \ri. \ri. } \\
   & & \le. \le. \cdot \le( 
   \frac{\partial}{\partial \bar{\Ps}_{\b ij}} \bar{\Ps}_{\a_{2n-1}} \ri)
   \Ps_{\b} \bar{\Ps}_{\a_1} \Ps_{\a_2} \cd \bar{\Ps}_{\a_{2n-1}} \ri)_{ij}
   \ri\rangle_S = - \ti{\ph} (n-1), \\
   \lefteqn{\lim_{N_m \rar \ift} \lim_{N_v \rar \ift}
   \frac{1}{N_m^2 \sqrt{N_v}} \sum_{i,j=1}^{N_m} \sum_{\b = 1}^{N_v}  
   \sum_{\a_1, \a_2, \ld, \a_{2n-1}}^{N_v} \le\langle \le(
   \bar{\Ps}_{\a_1} \Ps_{\a_2} \cd \bar{\Ps}_{\a_{2n-1}} \Ps_{\b} \ri. \ri. } 
   \\ & & \le. \le. \cdot \le( 
   \frac{\partial}{\partial \bar{\Ps}_{\b ij}} \bar{\Ps}_{\a_1} \ri)
   \Ps_{\a_2} \cd \bar{\Ps}_{\a_{2n-1}} \ri)_{ij} \ri\rangle_S = 
   \ti{\ph} (n-1), 
\eea
and
\bea
   \lefteqn{\lim_{N_m \rar \ift} \lim_{N_v \rar \ift}
   \frac{1}{N_m^2 \sqrt{N_v}} \sum_{i,j=1}^{N_m} \sum_{\b = 1}^{N_v}} \\ 
   & & \sum_{\a_1, \a_2, \ld, \a_{2n-1}}^{N_v} \le\langle \le(
   \bar{\Ps}_{\a_1} \Ps_{\a_2} \cd \bar{\Ps}_{\a_{2n-1}} \Ps_{\b} 
   \bar{\Ps}_{\a_1} \Ps_{\a_2} \cd \Ps_{\a_{2k}} \ri. \ri. \\
   & & \le. \le. \cdot \le( 
   \frac{\partial}{\partial \bar{\Ps}_{\b ij}} \bar{\Ps}_{\a_{2k+1}} \ri)
   \bar{\Ps}_{\a_{2k+2}} \cd \bar{\Ps}_{\a_{2n-1}} \ri)_{ij} 
   \ri\rangle_S = \ti{\ph} (n-k-1) G (k) 
\eea
for $1 \leq k \leq n-1$, it follows from Eq.(\ref{4.1.1}) that the
{\em Schwinger--Dyson equations} are given by
\beq
   - G(1) + c \ti{\ph} (1) + g G(2) = 0 
\la{4.1.2}
\eeq
and
\beq
   2 \sum_{k=0}^{n-2} \ti{\ph} (k) G(n-1-k) - G(n) + c \ti{\ph} (n) + g G(n+1) 
   = 0
\la{4.2}
\eeq
for $n$ = 2, 3, 4, \ld, and so on.

Let
\[ \ph(z) := \sum_{n=0}^{\ift} \ti{\ph} (n) z^{2n} \]
be a generating function of the full Green functions of the bosonic matrix 
$\ph^4$-theory.  It is well known \cite{bipz} that
\beq
   \ph(z) = \frac{1}{2z^2} - \frac{g}{2z^4} - \frac{1}{z^2}
   \le( \frac{1}{2} - g \g^2 - \frac{g}{2z^2} \ri) \sqrt{1 - 4 \g^2 z^2},
\la{4.3}
\eeq
where
\beq
   \g^2 := \frac{1- \sqrt{1 - 12g}}{6g}.
\la{4.3.1}
\eeq
Let
\[ \om_1 (z) := \sum_{n=0}^{\ift} G(n) z^{2n} \]
be a generating function of the full Green functions of the fermionic 
matrix-vector model.  It then follows from Eqs.(\ref{4.1.2}) and (\ref{4.2}) 
that
\beq
   \om_1 (z) = 1 + \frac{g G(1) + c - c \ph(z)}{\le\lb g - (1 - 2 g \g^2) z^2
   \ri\rb \sqrt{1 - 4 \g^2 z^2}} z^2.
\la{4.4}
\eeq
It is clear from Eq.(\ref{4.3}) that $\ph(z)$ has a branch cut between 
$-1/(2 \g)$ and $1/(2 \g)$ but is otherwise holomorphic on the complex 
$z$-plane.  The usual practice of matrix model analysis is to demand that the 
holomorphic structure of the generating function be as simple as possible 
\cite{wadia, david90, staudacher, douglasli, mz, ss}.  Moving along the same 
spirit, we assert that $\om_1 (z)$ is holomorphic on the whole complex plane
except the same branch cut.  In particular, the singularities at 
\[ z = \pm \sqrt{\frac{g}{1 - 2 g \g^2}} \]
are removable.  Then the numerator of the second term of the right hand side 
of Eq.(\ref{4.4}) vanishes if $z^2 = g / (1 - 2 g \g^2)$.  This fixes
$G(1)$:
\beq
   G(1) = \frac{c}{g} \ph \le( \sqrt{\frac{g}{1 - 2 g \g^2}} \ri) 
   - \frac{c}{g}.
\la{4.5}
\eeq
Let
\[ \om (z) := \frac{1}{z} \om_1 (\frac{1}{z}). \]
Then Eqs.(\ref{4.4}) and (\ref{4.5}) yield
\beq
   \om (z) = 1 - \frac{cz^2}{2} + \r (z), 
\la{4.5.1}
\eeq
where
\beq
   \r (z) := \frac{ cz \le\lb 2 (\g^2 + 2) - 3 z^2 + 3 g z^4 \ri\rb}
   {6 \le( g z^2 - 1 + 2 g \g^2 \ri) \sqrt{z^2 - 4 \g^2}}
\la{4.6}
\eeq
has a branch cut between $-2 \g$ and $2 \g$.  It is clear from Eq.(\ref{4.6})
that $G(n) = 0$ for any positive value of $n$ if $c = 0$.  This is consistent
with the well-known fact in fermionic matrix models that all even moments
vanish if the model respects chiral transformation \cite{mz}
\[ \Ps \rar \bar{\Ps} \; \mbox{and} \; \bar{\Ps} \rar - \Ps. \]

Criticality occurs if the zeros of $\r(z)$ coalesce with an end-point of the
branch cut.  It then follows from Eqs.(\ref{4.6}) and (\ref{4.3.1}) that the
critical values of $g$ and $z$ are
\[ g_* = 1/12 \; \mbox{and} \; z_* = 2 \sqrt{2}, \]
respectively.  Set
\[ g = g_* \exp (- a^2 \La_s) \; \mbox{and} \; z = z_* \exp (a \La_l), \]
where $a$ is the {\em cut-off length} and $\La_s$ and $\La_l$ are the
{\em continuum bulk and boundary cosmological constants}, respectively.
(c.f. Ref.\cite{david90}.)  At the {\em continuum limit}, $a$ is small and
Eq.(\ref{4.5.1}) becomes
\[ \om (z) \simeq 2c a^{-\frac{1}{2}} w(\La_s, \La_l), \]
where
\[ w(\La_s, \La_l) := \frac{1}{\le( \sqrt{\La_s} + 2 \La_l 
   \ri)^{\frac{1}{2}}} \]
may be regarded as the {\em renormalized tree-level one-boundary amplitude} of 
the orbifolded string worldsheet.  This is different from the corresponding 
amplitude of ordinary quantum gravity.  The scaling behavior of this fermionic 
matrix-vector model near criticality can be determined as usual by the 
{\em string susceptibility}
\beq
   \c & := & \lim_{N_m \rar \ift} \lim_{N_v \rar \ift}
   \frac{1}{N_m^4} \frac{\partial^2}{\partial g^2} \langle
   \log Z_{N_m, N_v} \rangle_S
\la{4.7.1} \\
   & = & \lim_{N_m \rar \ift} \lim_{N_v \rar \ift}
   \frac{1}{N_m^2} \frac{\partial}{\partial g} \le\langle
   \frac{1}{4} \sum_{\m_1, \m_2, \m_3, \m_4 = 1}^{N_v} \le\lb \Tr \le(
   \bar{\Ps}_{\m_1} \Ps_{\m_2} \bar{\Ps}_{\m_3} \Ps_{\m_4} \ri) \ri\rb^2
   \ri\rangle_S \nn \\
   & = & \frac{\partial}{\partial g} \le( \ti{\ph} (2) \ri).
\la{4.8}
\eeq
The {\em critical exponent} $\g_{\rm str}$ is defined by the formula
\[ \c \simeq (g - g_*)^{- \g_{\rm str}} \]
up to a proportionality constant which depends on $g_*$ only.  It then follows 
from Eqs.(\ref{4.8}), (\ref{4.3}), and (\ref{4.3.1}) that
\[ \g_{\rm str} = -1/2, \]
just like that of ordinary quantum gravity without matter \cite{dgz}.

\section{Another Quantum Orbifold Geometry}
\la{s5}

We may arrange the fermionic matrix-vectors differently to form different
models.  For instance, consider another family of fermionic matrix-vector
models whose actions take the form
\beq
   \lefteqn{S' := N_m \sqrt{N_v} \sum_{\m = 1}^{N_v}
   \Tr \bar{\Ps}_{\m} \Ps_{\m}} \nn \\
   & & + N_m \sum_{n=1}^{\ift} c'_n
   \sum_{\m_1, \m_2, \ld, \m_{2n} = 1}^{N_v} \Tr \le(
   \bar{\Ps}_{\m_1} \Ps_{\m_2} \bar{\Ps}_{\m_3} \Ps_{\m_4} \cd
   \bar{\Ps}_{\m_{2n-1}} \Ps_{\m_{2n}} \ri. \nn \\
   & & \le. \cdot \Ps_{\m_{2n}} \bar{\Ps}_{\m_{2n-1}}
   \Ps_{\m_{2n-2}} \bar{\Ps}_{\m_{2n-3}} \cd \Ps_{\m_2} \bar{\Ps}_{\m_1} \ri) 
   \nn \\
   & & + N_m^2 \sum_{n=2}^{\ift} \frac{g'_n}{2n}
   \sum_{\m_1, \m_2, \ld, \m_{2n} = 1}^{N_v} \Tr \le(
   \bar{\Ps}_{\m_1} \Ps_{\m_2} \bar{\Ps}_{\m_3} \Ps_{\m_4} \cd
   \bar{\Ps}_{\m_{2n-1}} \Ps_{\m_{2n}} \ri) \nn \\
   & & \cdot \Tr \le( \Ps_{\m_{2n}} \bar{\Ps}_{\m_{2n-1}}
   \Ps_{\m_{2n-2}} \bar{\Ps}_{\m_{2n-3}} \cd \Ps_{\m_2} \bar{\Ps}_{\m_1} \ri).
\la{5.1}
\eeq
The partition function $Z'_{N_m, N_v}$ is given by Eq.(\ref{3.2}) with $S$
replaced with $S'$.  We are interested in the full Green functions which read
\bea
   G'_{N_m, N_v} (n) & := & \frac{1}{N_m}
   \sum_{\m_1, \m_2, \ld, \m_{2n} = 1}^{N_v} \langle \Tr \le(
   \bar{\Ps}_{\m_1} \Ps_{\m_2} \bar{\Ps}_{\m_3} \Ps_{\m_4} \cd
   \bar{\Ps}_{\m_{2n-1}} \Ps_{\m_{2n}} \ri. \\
   & & \le. \cdot \Ps_{\m_{2n}} \bar{\Ps}_{\m_{2n-1}}
   \Ps_{\m_{2n-2}} \bar{\Ps}_{\m_{2n-3}} \cd \Ps_{\m_2} \bar{\Ps}_{\m_1}
   \ri) \rangle_{S'}.
\eea
The only changes to the Feynman rules are in the vertices.  Some vertices are 
depicted in Fig.\ref{f6}.

\begin{figure}
\epsfxsize=4in
\centerline{\epsfbox{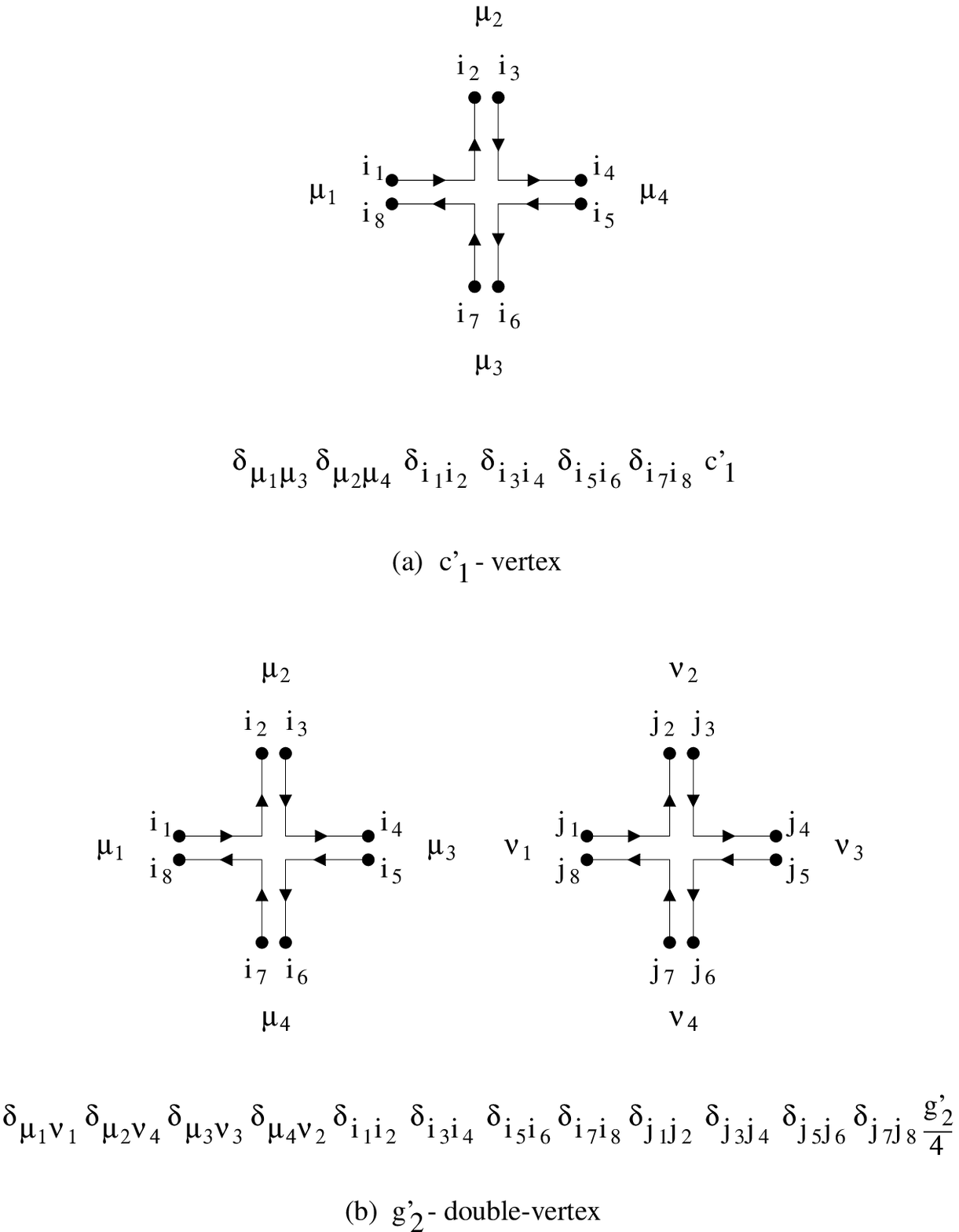}}
\caption{\em Selected vertices of the fermionic matrix-vector model defined
in Eq.(\ref{5.1}).  Depicted are (a) a vertex with a coefficient $c'_1$ and
(c) a double-vertex with a coefficient $g'_2$.}
\la{f6}
\end{figure}

The double large-$N$ limit is taken as in Eq.(\ref{3.3}) with $Z$,
$Z_{N_m, N_v}$, $G(n)$, and $G_{N_m, N_v} (n)$ replaced with their primed
versions.  A Feynman diagram which survives the double large-$N$ limit is
illustrated in Fig.\ref{f7}.  If we choose the center of the great circle to
be the origin of a two-dimensional Cartesian coordinate system and point the
$y$-axis to $\m_1$, then the diagram together with its dual respects 
reflection about the $y$-axis.  This suggests that {\em the dual of the 
Feynman diagrams of $G'(n)$ in the double large-$N$ limit may be conceived of 
as discrete random surfaces which respect reflection about the $y$-axis, i.e., 
these are random surfaces of a bounded region of the orbifold ${\bf R}^2 / 
{\bf Z}_2$}.  (c.f. Fig.\ref{f8}.)

\begin{figure}[ht]
\epsfxsize=4in
\centerline{\epsfbox{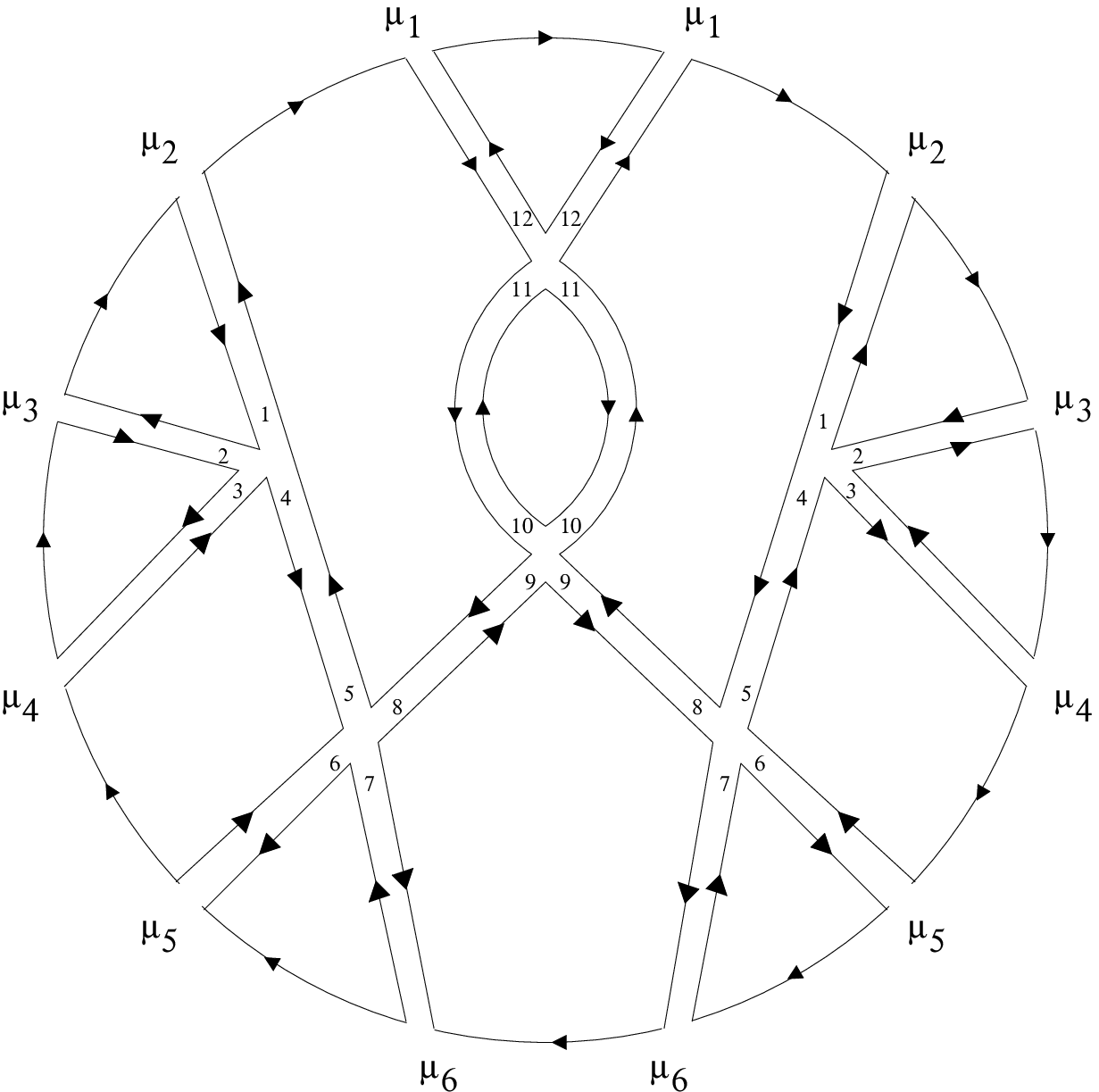}}
\caption{\em A Feynman diagram of $G'(3)$.  Only vector indices are shown.  
The numbers 1, 2, 3, \ld, and so on stand for $\n_1$, $\n_2$, $\n_3$, and so 
on, respectively.}
\la{f7}
\end{figure}

\begin{figure}[ht]
\epsfxsize=4in
\center{\epsfbox{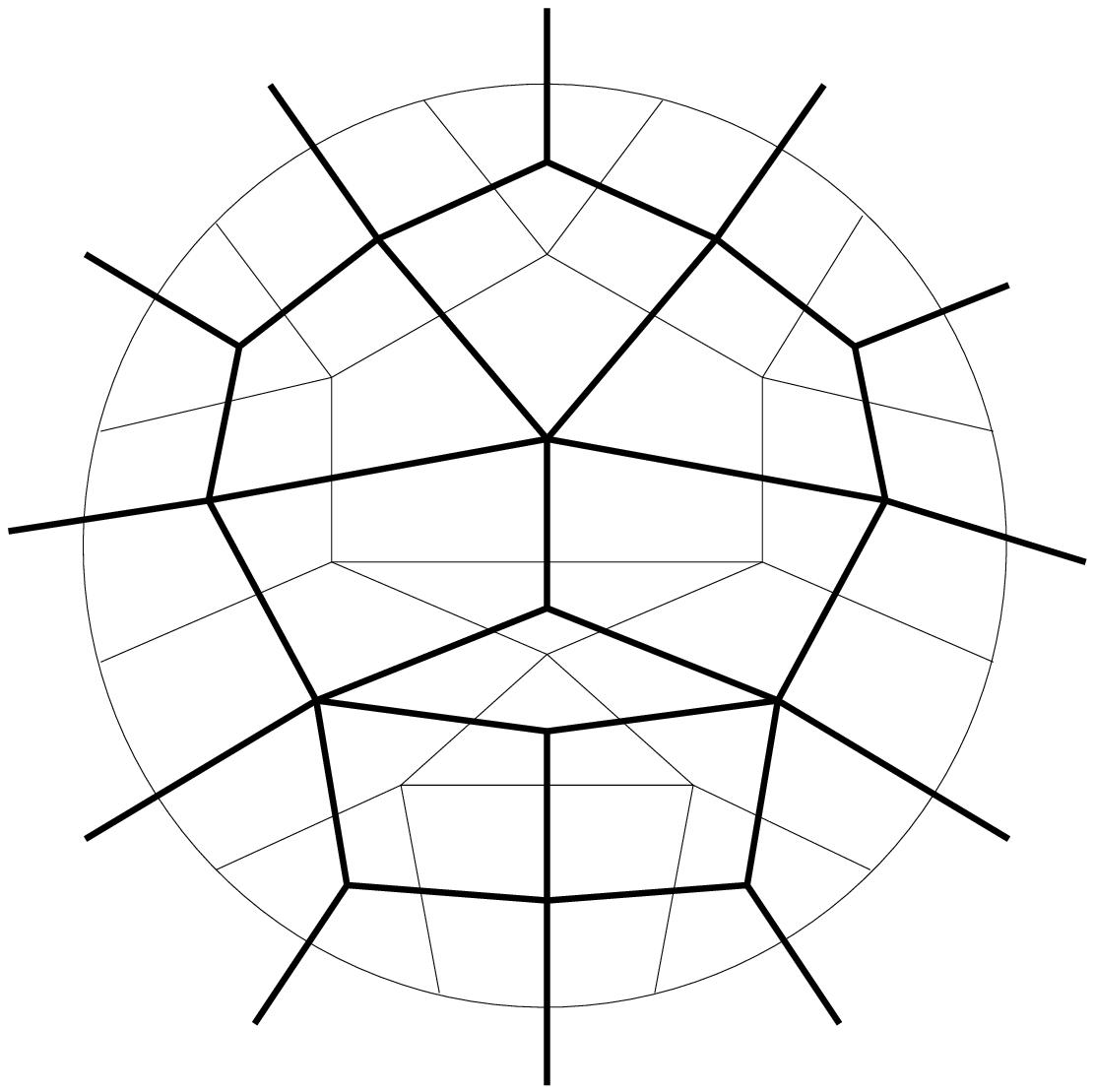}}
\caption{\em A discretized random orbifold surface with a boundary generated
by the fermionic matrix-vector model defined in Eq.(\ref{5.2}).  This surface 
respects reflection symmetry instead of parity.  the ``spokes'' protruding 
from the great circle converge to the same point on a spherical surface.}
\la{f8}
\end{figure}

Let us concentrate on the simplest model in this family and choose only the 
coefficients $c = c'_1$ and $g = g'_2$ to be non-zero.  It follows from 
Eq.(\ref{5.1}) that
\beq
   S' & := & N_m \sqrt{N_v} \sum_{\m = 1}^{N_v} \Tr \bar{\Ps}_{\m} \Ps_{\m}
   + N_m c \sum_{\m_1, \m_2 = 1}^{N_v} \Tr \le(
   \bar{\Ps}_{\m_1} \Ps_{\m_2} \Ps_{\m_2} \bar{\Ps}_{\m_1} \ri) \nn \\
   & & + \frac{N_m^2 g}{4} \sum_{\m_1, \m_2, \m_3, \m_4 = 1}^{N_v}
   \Tr \le( \bar{\Ps}_{\m_1} \Ps_{\m_2} \bar{\Ps}_{\m_3} \Ps_{\m_4} \ri)
   \Tr \le( \Ps_{\m_4} \bar{\Ps}_{\m_3} \Ps_{\m_2} \bar{\Ps}_{\m_1} \ri).
\la{5.2}
\eeq
To obtain the full Green functions, consider the trivial equation
\bea
   \lefteqn{\lim_{N_m \rar \ift} \lim_{N_v \rar \ift}
   \frac{1}{N_m^2 \sqrt{N_v} Z'_{N_m, N_v}} \sum_{i,j=1}^{N_m} 
   \sum_{\b = 1}^{N_v} \int d\Ps_1 d\bar{\Ps}_1 d\Ps_2 d\bar{\Ps}_2 \cd
   d\Ps_{N_v} d\bar{\Ps}_{N_v} } \nn \\
   & & \frac{\partial}{\partial \bar{\Ps}_{\b ij}} \{
   \sum_{\a_1, \a_2, \ld, \a_{2n-1}}^{N_v} \le( \bar{\Ps}_{\a_{2n-1}}
   \Ps_{\a_{2n-2}} \cd \bar{\Ps}_{\a_1} \bar{\Ps}_{\a_1} \Ps_{\a_2}
   \cd \bar{\Ps}_{\a_{2n-1}} \Ps_{\b} \ri)_{ij} \nn \\
   & & \exp S' \} = 0
\eea
for any positive integer $n$.  After some manipulations, this equation leads
to the Schwinger--Dyson equation
\beq
   \sum_{k=1}^n \ti{\ph} (n-k) G'(k-1) + (1 + c) G'(n) + g G'(n+1) = 0,
\la{5.3}
\eeq
where $n$ is any positive integer and $\ti{\ph}(n)$ was defined in 
Eq.(\ref{2.3}) again with 
\[ V(M) = \frac{1}{2} M^2 - \frac{g}{4} M^4. \]
Introduce the following generating function of the moments:
\[ \om'_1 (z) := \sum_{n=0}^{\ift} G'(n) z^{2n}. \]
Then we may rewrite Eq.(\ref{5.3}) as
\[ \om'_1 (z) = \frac{g + \le\lb 1 + c + g G'(1) \ri\rb z^2}
   {g + (1 + c) z^2 + z^4 \ph(z)}, \]
where $\ph(z)$ was defined in Eqs.(\ref{4.3}) and (\ref{4.3.1}).  Hence,
\beq
   \om'_1 (z) = \frac{ P_1 (z^2) \le\lb \frac{g}{2} +
   \le( \frac{3}{2} + c \ri) z^2 - \ti{\r} (z^2) \ri\rb }{P_3 (z^2)},
\la{5.4}
\eeq
where
\bea
   P_1 (\z) & := & g + \le\lb 1 + c + g G'(1) \ri\rb \z, \\
   \ti{\r} (\z) & := & \le\lb \frac{g}{2} +
   \le( g \g^2 - \frac{1}{2} \ri) \z \ri\rb \sqrt{1 - 4 \g^2 \z},
\eea
and
\[ P_3 (\z) := \z \le\lb (2 + c) g + (2 + 3c + c^2 - g) \z +
   \le( -\frac{4}{3} g \g^2 + \frac{1}{9} \g^2 + \frac{8}{9} \ri) \z^2
   \ri\rb . \]
Zero and one of the non-zero roots of the cubic polynomial $P_3 (\z)$ are
roots of $\ti{\r} (\z)$, too.  Hence $G'(1)$ is fixed by the requirement
that $P_1 (\z)$ vanish if $\z$ is equal to the other non-zero root of
$P_3 (\z)$.  Then we may read off the full Green functions from $\om'_1 (z)$.

At the critical point, the zeros of
\[ \r' (z) := - \frac{ P_1 (z^2) \ti{\r} (z^2) z}{P_3 (z^2)} \]
coalesce with an end-point of its branch cut.  Thus the critical values of 
$g$, $c$, and $z$ are
\[ g_* = \frac{1}{12}, \; c_* = - \frac{11}{6}, \; \mbox{and} \;
   z_* = \frac{1}{2 \sqrt{2}}. \]
Near the critical point,
\[ g = g_* \exp (- a^2 \La_s), \; c = c_* \exp (- a^2 \La_c), \; \mbox{and}
   z = z_* \exp (- a \La_l), \]
where $a$ is small at the continuum limit.  Then
\[ \r' (z) \simeq - \frac{1}{4 \sqrt{2}} w'(\La_s, \La_l) \sqrt{a}, \]
where
\[ w' (\La_s, \La_l) := \frac{1}{\La_l} \le( \sqrt{\La_s} - 4 \La_l \ri)
   \le( \sqrt{\La_s} + 2 \La_l \ri)^{\frac{1}{2}} \]
is the renormalised tree-level one-boundary amplitude of this orbifolded 
string worldsheet.  The string susceptibility is defined as in 
Eq.(\ref{4.7.1}) with $S$ replaced with $S'$.  It turns out that
\[ \c = \frac{\partial}{\partial g} \le( \ti{\ph} (2) \ri)
   \simeq (g - g_*)^{\frac{1}{2}}. \]
Hence we conclude again from Eqs.(\ref{4.3}) and (\ref{4.3.1}) that
\[ \g_{\rm str} = -1/2. \]
The critical exponent is identical to but the tree-level one-boundary 
amplitude is different from that of ordinary quantum gravity.

\section{Conclusion and Outlook}
\la{s6}

It is clear from the above calculation that fermionic matrix-vector models
may be used to analyse the behavior of different quantum orbifold geometry.
They have the same critical exponent as that of ordinary quantum gravity, but 
the one-boundary amplitudes are already different at the tree level.  It is 
clearly of interest to study the multi-loop amplitudes and double-scaling 
or even triple-scaling limit of these models as well as their multicritical 
generalisations and explore the subsequent ramifications in quantum gravity 
and string theory.  

On the more mathematical side, clarification of the relationship between the 
set of fermionic matrix-vectors of infinite order with the ordered pair of a 
C$^*$ algebra and a vector space, constituents of a noncommutative probability
theory of type~B as defined in Ref.\cite{bgn}, will be desirable.  Fermionic 
matrix-vector models may even be useful in the applications of noncommutative 
probability of type~B to algebraic analysis.  (c.f. the application of 
ordinary random matrix models, the embodiment of noncommutative probability of 
type~A, to the analysis of von Neumann algebras \cite{voiculescu}.)  
Furthermore, just like it is interesting to construct rigorous topological or 
combinatorial arguments to relate the Feynman diagrams of ordinary matrix 
models with quadrangulated random surfaces \cite{biz, jackson}, it would be 
desirable to develop a rigorous proof for the identification of the dual of 
Feynman diagrams of fermionic matrix-vector models with quadrangulated 
orbifold surfaces and explore the consequences.

\vskip 1pc
\noindent \Large{\bf \hskip .2pc Acknowledgment}
\vskip 1pc

\normalsize

\noindent
I am grateful to A. Nica and S. G. Rajeev for providing me valuable insights
into the subject matter of this article.  I would also like to thank the 
referee for helpful suggestions for the revision of this article.

\end{document}